\documentstyle[preprint,eqsecnum,aps,twoside]{revtex}
\newcommand{\C}{{\if mm {{\rm C}\mkern -15mu{\phantom{\rm t}\vrule}}
\mkern +10mu \else \leavemode \hbox{I}\kern -.17em \hbox{C} \fi}}
\newcommand {\bJ}{\mbox{\bf J}} 
 
\newcommand {\bj}{\mbox{\bf j}} 
\newcommand {\br}{\mbox{\bf r}}

\newcommand {\bA}{\mbox{\bf A}}
\newcommand {\bB}{\mbox{\bf B}}
\newcommand {\bP}{\mbox{\bf P}}
\newcommand {\bM}{\mbox{\bf M}}

\newcommand {\bD}{\mbox{\bf D}}
\newcommand {\bH}{\mbox{\bf H}}
\newcommand {\bE}{\mbox{\bf E}}
\newcommand {\bq}{\mbox{\bf q}}

\newcommand {\bpi}{\mbox{\boldmath $\Pi$}}

\newcommand {\bT}{\mbox{\bf T}}
\newcommand {\bal}{\mbox{\boldmath $\alpha$}}
\newcommand {\bbe}{\mbox{\boldmath $\beta$}}
\newcommand {\bde}{\mbox{\boldmath $\delta$}}

\newcommand {\bp}{\mbox{\bf p}}
\newcommand {\p}{\partial}

\newcommand {\bC}{\mbox{\bf C}}

\newcommand {\vf}{\varphi}

\begin{document}
\title{Electrodymanics with Toroid Polarization}
\author{V.M.~Dubovik and B.~Saha \\
{\small \it Bogoliubov Laboratory of Theoretical Physics }\\
{\small \it Joint Institute for Nuclear Research}\\
{\small \it 141980, Dubna, Moscow reg., Russia\\
e-mail: dubovik@thsun1.jinr.dubna.su\\  
e-mail: saha@thsun1.jinr.dubna.su}}
\maketitle
\begin{abstract}
A modified system of equations of electrodynamics has been obtained.
Beside the Lagrangian one an alternative gauge-like formalism has been
developed to introduce the toroid moment contributions in the equations
obtained. The two potential formalism that was worked out by us earlier
has been developed further where along with the two vector potentials
we introduce two scalar potentials thus taking into account all the 
four equations of electromagnetism. 
\end{abstract}
 Maxwell theory: general mathematical aspects
\vskip 5mm
\section{Introduction} 
\label{sec:level1}
The history of electromagnetism is the history of struggle of different
rival concepts from the very early days of its existence. Though, after
the historical observation by Hertz, all main investigations in
electromagnetism were based on Maxwell equations, nevertheless this
theory still suffers from some shortcomings inherent to its predecessors.
Several attempts were made to remove the internal inconsistencies of the 
theory. To be short we refer to very few of them. One of the attempts to
modify the theory of electromanetism was connected with the introduction
of magnetic charge in Maxwell equation by Dirac~\cite{Dirac,Dirac1}, 
while keeping the usual definition of $\bE$ and $\bB$ in terms of the gauge 
potentials. Recently D. Singleton~\cite{Singleton} 
developed this theory introducing two four-vector potentials 
$A^\mu = (\phi_e, \bA)$ and $C^\mu = (\phi_m, \bC)$. 
Note that, a similar theory (two potential formalism) was
developed by us few years ago (we will come back to it in Sec. 3). The
main defect of the theory developed by Singleton in our view is that
the existence of magnetic charge still lack of experimental support,
hence can be considered as a mathematically convenient one only.
Recently Chubykalo a.o. made an effort to modify the electromagnetic 
theory by invoking both the transverse and longitudinal (explicitly 
time independent) fields simultaneously, thus giving an equal footing 
to both the Maxwell-Hertz and Maxwell-Lorentz equations
~\cite{Chubykalo}. To remove all ambiguities related to the 
applications of Maxwell's displacement current they substituted all partial 
derivatives in Maxwell-Lorenz equations by the {\it total} ones and separated
all field quantities into {\it two independent classes} with explicit 
$\{\}^*$ and implicit $\{\}_0$ time dependence, respectively. 
Another attempt to modify the equations of eletromagnetism is connected
with the existence of the third family of multipole moments, namely
the {\it toroid} one. This theory was developed by us during the recent
years. Recently we introduced toroid moments in Maxwell equations exploiting
Lagrangian formalism~\cite{ICTP}. In the Sec. 2 of this paper we give a brief
description of this formalism. Moreover, here we develop an alternative
method to introduce toroid moments in the equation of electromagnetism.
In Sec. 3 we develop two potential formalism suggested by us earlier. 
\section{Introduction of toroid moments in the equations of electromagnetism} 
\label{sec:level2}
\vskip 3mm
Ya.~Zel'dovich~\cite{Zel'dovich} was the first to
introduce {\it anapole} in connection with the global electromagnetic
properties of a toroid coil that are impossible to describe within the
charge or magnetic dipole moments in spite of explicit axial symmetry
of the toroid coil. Further, in 1974 Dubovik and Cheskov~\cite{DC} 
determined the toroid moment in the framework of classical electrodynamics.
Recently a principally new type of magnetism known as {\it aromagnetism} was 
observed in a class of organic substances, suspended either in water or in 
other liquids~\cite{Spartakov}. Later, it was shown that this phenomena
of aromagnetism cannot be explained in a standard way, e.g., by 
ferromegnetism, since the organic molecules do not possess magnetic moments
of either orbital or spin origin. It was also shown that the origin of
aromagnetism is the interaction of vortex electric field induced by 
alternative magnetic one with the axial toroid moments in aromatic elements
~\cite{MM}. 
These experimental results force the introduction of toroid 
moments in the framework of conventional classical electrodynamics that in 
its part inevitably leads to the modification of the equations
of electromagnetism. In the two following subsection we give two
alternative schemes of introduction of toroid moments in the electromagnetic
equations.
\subsection{Lagrangian Formalism}
As a starting point we consider the interacting system of electromagnetic 
field and non-relativistic charged particles given by the Lagrangian 
density~\cite{Healy}
\begin{eqnarray} 
\label{L1}
L&=& L_{\rm par}\,+\,L_{\rm rad}\,+\,L_{\rm int} \\
L_{\rm par}&=&\frac{1}{2}\,\sum_{\alpha}\, 
m_{\alpha}\,\dot{\bq}_{\alpha}^{2} \,-\,\frac{1}{2}\,\sum_{\alpha \ne 
\beta}\frac{e_{\alpha}\,e_{\beta}} {|{\bq}_{\alpha}\,-\,{\bq}_{\beta}|} 
\nonumber \\
L_{\rm rad}&=&
\frac{1}{8\pi}\, \int\, \bigl[\frac{\dot{\bA}      
^2}{c^2}\,- \,(\mbox{curl}\,{\bA})^2\bigr]\, d\br \nonumber \\
L_{\rm int}&=&\frac{1}{c}\,\int\,{\bJ}({\br}) \cdot {\bA}({\br})\, d\br
\,=\,\sum_{\alpha}\frac{e_{\alpha}}{c}\,\dot{\bq}_{\alpha} \cdot 
{\bA}(q_\alpha, t). \nonumber
\end{eqnarray}
Here $L_{\rm par}$ is the Lagrangian appropriate to a system of charged 
particles interacting solely through instantaneous Coulomb force; it has 
the simple form of "kinetic energy minus potential energy". $L_{\rm rad}$ is 
the Lagrangian for a radiation field far removed from the charges and 
currents, and has the form of "electric field energy minus magnetic field 
energy". The interaction Lagrangian $L_{\rm int}$ couples the particle 
variables to the field ones. It can be easily verified that variation 
with respect to the particle coordinates gives the second law of Newton 
with the Lorentz force  
\begin{equation}
\label{LF} 
m_\alpha\,\ddot{\bq}_\alpha\,=\,e_\alpha\,{\bE}({\bq}_\alpha,\,t)\,+\,
\frac{e_\alpha}{c}\dot{\bq}_\alpha \times {\bB}({\bq}_\alpha,\,t).
\end{equation}
Variation of the Lagragian~(\ref{L1}) with respect to field variables
gives the equation of motion for the vector potential
\begin{equation}
{\rm curl\,curl}\,{\bA}\,+\,\frac{1}{c^2}\,\frac{\p^2 \bA}{\p t^2}\,
=\,-\,\frac{4\pi}{c}\,{\bJ}
\end{equation}
Defining $\bB = {\rm curl}\ \bA$ and $\bE = - \dot{\bA}/c$ one obtains
\begin{equation}
{\rm curl}\ \bB = \frac{1}{c} \frac{\p \bE}{\p t} + \frac{4 \pi}{c} \bJ
\label{CE}
\end{equation}
It should be emphasized that $\bE$ in ~(\ref{LF}) and ~(\ref{CE}) is
the transverse part of the total electric field. The longitudinal electric
field in question is entirely electrostatic.
The Hamiltonian, corresponding to the Lagrangian~(\ref{L1}) reads              
\begin{eqnarray}
H[{\bpi}, {\bA}; p,q] \,&=&\, \sum_{\alpha}\,{\bp}_\alpha\cdot 
\dot{\bq}_\alpha \,+\,\int\,{\bpi}\cdot \dot{\bA} d\br\,-\, L \nonumber \\
&=& \sum_{\alpha}\frac{1}{2m_\alpha} 
\bigl[{\bp}_\alpha\,-\,\frac{e_\alpha}{c}\,{\bA}({\bq},\,t) \bigr]^2\,+\,
\frac{1}{2}\,\sum_{\alpha \ne \beta}\frac{e_{\alpha}\,e_{\beta}}
{|{\bq}_{\alpha}\,-\,{\bq}_{\beta}|}  \\
& &+ \frac{1}{8\pi}\, \int\, \bigl[(4\pi\,c {\bpi})^2 
\,+ \,(\mbox{curl}\,{\bA})^2\bigr]\, d\br, \nonumber
\end{eqnarray} 
where the corresponding conjugate momenta are  
$${\bp}_\alpha\,=\,
m\,\dot{\bq}_\alpha\,+\,(e_\alpha/c)\,{\bA}({\bq},\,t), \quad
{\bpi}({\br})\,=\,(4\pi c^2)^{-1} \dot{\bA}.$$
It is well known that in classical dynamics the addition of a total
time derivative to a Lagrangian leads to a new Lagrangian with the 
equations of motion unaltered. Lagrangians obtained in this manner are 
said to be equivalent. In general, the Hamiltonians following from the 
equivalent Lagrangians are different. Even the relationship between the 
conjugate and the kinetic momenta may be changed~\cite{PT}.
Moreover, let us notice that the basic equations of any new theory cannot 
be introduced strictly deductively. Usually, either they are postulated 
in differential form based on the partial integral conservation laws or 
transformations of basic dynamical variables, whose initial definitions 
usually have some analog in mechanics. Let us remark that we need to do 
so not only by inertia of thinking but also because of the fact that 
most of our measurements have its objects as individual particles or use 
them as test one. The situation is the same in electromagnetism and in 
gravitation. In general geometrical interpretation of dynamical 
variables plays the crucial role. 
An equivalent Lagrangian to that of~(\ref{L1}) is~\cite{ICTP}  
\begin{equation} 
L^{\rm equiv}\,=\,L\,-\,\frac{1}{c}\frac{d}{dt}\bigl[\int\,\{{\bP}({\br})\,
+\, \mbox{curl}{\bT}^e({\br})\}\cdot{\bA}({\br})\,dV\bigr],
\end{equation}
where the toroid contribution has been taken into account.
Here $\bT^{e}$ is axial toroid moment (ATM) is electrical by nature
(toroid dipole polarization vector of electric type). 
Writing it in the explicit form we get
The field conjugate to the vector potential $\bA$ is now
\begin{eqnarray}
4 \pi c {\bpi}({\br}) := - \bD (\br) = 
- \bigl({\bE}({\br}) + 4\pi ({\bP}({\br})
+ \mbox{curl}{\bT}^e({\br}))\bigr) \nonumber 
\end{eqnarray} 
Since only the free field $\bE$ is generated due to the change of 
magnetic field $\bB$ one writes  
\begin{equation}  \label{D1}  
\mbox{curl}{\bD}({\br}) = -\frac{1}{c} \dot{\bB}({\br}) +
4\pi (\mbox{curl}{\bP}({\br}) + \mbox{curl\,curl}{\bT}^e({\br})),  
\end{equation}          
under  $\mbox{curl}{\bE}({\br}) = - \dot{\bB}({\br})/c$.  
The new Lagrangian is a function of the variables ${\bq}_\alpha$,
$\dot{\bq}_\alpha$ and a functional of the field variables ${\bA}$, 
$\dot{\bA}$, and the equations of motion follow from the variational 
principle. Applying the Euler-Lagrange equations of motion one gets
~\cite{ICTP}
\begin{equation} \label{B1}
\mbox{curl}{\bB}({\br}) = \frac{1}{c}\dot{\bD}({\br}) + 
\frac{4 \pi}{c} \bj_{\rm free} + 4\pi
\bigl(\mbox{curl}{\bM}({\br}) +\mbox{curl\,curl}{\bT}^m({\br})\bigr)
\end{equation}
Here the currents were divided into free and bound state (due to electric 
polarization and magnetization) one as~\cite{Purcell}
\begin{eqnarray}
{\bJ}({\br}) = \bj_{\rm free} +
c\,\mbox{curl}{\bM}({\br}) + \dot{\bP}({\br}) 
\end{eqnarray}
and an additional condition on $\bT^e$ is imposed
\begin{equation}
{\rm curl}\, \bT^{m,e} = \pm \frac{1}{c} 
\dot{\bT}^{e,m}
\label{AC}
\end{equation}
where $\bT^m$ is the toroid dipole polarization vector of magnetic type.
The relation~(\ref{AC}) demands some comments. Both $\bT^e$ and $\bT^m$ 
represent the closed isolated lines of electric and magnetic fields. So 
they have to obey the usual differential relations similar to the free 
Maxwell equations~\cite{DK,Miller}). However, 
remark that signs here are opposite to the corresponding one in
Maxwell equations because the direction of electric dipole 
is accepted to be chosen opposite to its inner electric field~\cite{Ginzburg}.
If we define the auxiliary field $\bH$ to be 
\begin{equation} 
{\bH}({\br})\,=\,{\bB}({\br})\,-\,4\pi\, 
({\bM}({\br})\,+\,\mbox{curl}{\bT}^m({\br})) 
\end{equation} 
then it deduces to 
$$ \mbox{curl}{\bH}\,=\,\frac{1}{c}\dot{\bD} + 
\frac{4 \pi}{c}\,\bj_{\rm free} $$  
But the latter formula is unsatisfactory from the physical point of view.  
It is easy to image the situation when $\bB$ and $\bM$ are absent, 
because the medium may be composed from isolated aligned dipoles $\bT^m$ 
~\cite{Zhel,DMM,DLM} and each $\bT^m$ is the source of 
free-field (transverse-longitudinal) potential but not $\bB$ \cite{DT}. 
So the transition to the description by means of potentials is inevitable.
The Hamiltonian, corresponding to the equivalent Lagrangian in this case 
reads
\begin{eqnarray}
H^{\rm equiv}[{\bpi}, {\bA}; p,q] &=& 
\sum_{\alpha}\frac{1}{2m_\alpha} 
\bigl[{\bp}_\alpha - \frac{e_\alpha}{c} {\bA}({\bq}, t) \bigr]^2 +
\frac{1}{2}\sum_{\alpha \ne \beta}\frac{e_{\alpha}\,e_{\beta}}
{|{\bq}_{\alpha} - {\bq}_{\beta}|} \nonumber \\
& &+ \frac{1}{8\pi} \int \bigl\{[4\pi(\bP + \mbox{curl}\,\bT^e)-\bD]^2 
+ (\mbox{curl}\,{\bA})^2\bigr\} d\br \\
& & +\frac{1}{c}\int \bJ\cdot \bA d\br - \int 
{\bM}\cdot{\bB} d\br - \int \bB\cdot\mbox{curl}\,{\bT}^m d\br.\nonumber  
\end{eqnarray} 
\subsection{Gauge-like Transformation}
The Maxwell equations for electromagnetic fields in media can be written as
\begin{mathletters}
\label{Eq:al1}
\begin{eqnarray}
{\rm curl} \bH - \frac{1}{c}\frac{\p \bD}{\p t} &=& 
\frac{4 \pi}{c} \bj_{\rm free}  \label{E1a}  \\  
{\rm div} \bD &=& 4 \pi \rho  \label{E1b}\\
{\rm curl} \bE + \frac{1}{c}\frac{\p \bB}{\p t} &=& 0 \label{E1c}  \\
{\rm div} \bB &=& 0  \label{E1d} 
\end{eqnarray}
\end{mathletters}
where 
\begin{mathletters}
\begin{eqnarray}
\bD &=& \bE + 4 \pi \bP  \\
\bH &=& \bB - 4 \pi \bM 
\end{eqnarray}
\end{mathletters}
In the previous subsection we introduced toroid moments into Maxwell 
equation through Lagrangian formalism. In doing so we first constructed 
an equivalent Lagrangian. Here we do the same using in an alternative way
which looks rather a gauge transformation. To this end we introduce
two vectors $\bT^m$ and $\bT^e$ (toroid dipole polarization vector
of magnetic type and toroid dipole polarization of electric type,
respectively) such that 
\begin{mathletters}
\label{Eq:al3}
\begin{eqnarray}
\bP &\Longrightarrow& \bP + {\rm curl} \bT^e \label{E3a}\\
\bM &\Longrightarrow& \bM + {\rm curl} \bT^m \label{E3b}
\end{eqnarray}
\end{mathletters}
It can be easily shown that the system~(\ref{Eq:al1}) is invariant under 
the transformation~(\ref{Eq:al3}) if we impose the additional 
condition~(\ref{AC}), i.e.,
\begin{equation}
{\rm curl} \bT^{e,m} = \pm \frac{1}{c}\frac{\p \bT^{m,e}}{\p t} \label{E5}
\end{equation}
In account of~(\ref{Eq:al3}) and~(\ref{E5}) we rewrite the system~(\ref{Eq:al1}) 
as
\begin{mathletters}
\label{Eq:al6}
\begin{eqnarray}
{\rm curl} \bB &=& \frac{1}{c}\frac{\p \bD}{\p t} + \frac{4 \pi}{c} 
\bj_{\rm free} + 4 \pi\{ {\rm curl}\,\bM + {\rm curl\,curl}\,\bT^m \}
\label{E6a}  \\  
{\rm div} \bD &=& 4 \pi \rho  \label{E6b}\\
{\rm curl} \bD &=& - \frac{1}{c}\frac{\p \bB}{\p t} 
+ 4 \pi\{ {\rm curl}\,\bP + {\rm curl\,curl}\,\bT^e\}  \label{E6c}  \\
{\rm div} \bB &=& 0  \label{E6d} 
\end{eqnarray}
\end{mathletters}
As is seen the equations~(\ref{E6a}) and~(\ref{E6c}) of the system
~(\ref{Eq:al6}) completely coincide with~(\ref{B1}) and~(\ref{D1}) of the
previous subsection. Thus we introduced toroid moments in Maxwell equations
using two different formalisms.
\section{Two Potential Formalism}
\label{sec:level3}
It is commonly believed that the divergence equations of the Maxwell
system are "redundant". Recently Krivsky a.o.~\cite{Krivsky} 
claimed that to describe the free electromagnetic field it is sufficient 
to consider the curl-subsystem of Maxwell equations since the equalities 
${\rm div} \bE = 0$ and ${\rm div} \bB = 0$ are fulfilled identically. 
Contrary to this statement, Jiang and Co~\cite{Jiang} proved that the 
divergence equations are not redundant and that neglecting these equations 
is at the origin of spurious solutions in computational electromagnetics. 
Here we construct 
generalized formulation of Maxwell equations including both curl and 
divergence subsystems. In this section we develop two potential formalism 
(a similar formalism was developed by us earlier with the curl-subsystem 
taken into account only). Note that in the ordinary one potential formalism 
($\bA, \vf$) the second set of Maxwell equations are fulfilled 
identically. So that all the four Maxwell equations bring their contribution 
individually, in our view, one has to rewrite the Maxwell equation in terms 
of two vector and two scalar potentials. 
Because of introduction of toroid moments (see Sec. 2) now $\bB$ and $\bD$ 
have lost their initial meaning, hence should be reinterpreted. It means 
the deduction of the equation of evolution by inserting $\bB = 
{\rm curl} \bA$ and $\bE = -\dot{\bA}/c$ is valid no longer  
and we have to introduce some new potential that could explain the new
$\bB$ and $\bD$. To this end we introduce so-called double potential
~\cite{DMag,Kluwer,ICTP}. As was mentioned, due to
introduction of toroid moments the vectors $\bB$ and $\bD$ should be
redefined. We denote these new quantities as $\bbe$ and $\bde$, respectively.
In account of it, the system~(\ref{Eq:al6}) should be rewritten as
\begin{mathletters}
\label{New}
\begin{eqnarray}
{\rm curl} \bbe &=& \frac{1}{c}\frac{\p \bde}{\p t} + \frac{4 \pi}{c} 
\bj_{\rm free} + 4 \pi\{ {\rm curl}\,\bM + {\rm curl\,curl}\,\bT^m \} 
\label{N1}\\  
{\rm div} \bde &=& 4 \pi \rho \label{N2} \\
{\rm curl} \bde &=& - \frac{1}{c}\frac{\p \bbe}{\p t} 
+ 4 \pi\{ {\rm curl}\,\bP + {\rm curl\,curl}\,\bT^e\} \label{N3}   \\
{\rm div} \bbe &=& 0 \label{N4}  
\end{eqnarray}
\end{mathletters}
Before developing the two potential formalism we first rewrite 
system~(\ref{Eq:al1}) in terms of vector and scalar potentials $\bA, \phi$ 
such that 
$\bB = {\rm curl}\,\bA$, $\bE = -\nabla \vf - (1/c)(\p \bA/ \p t)$.
Following any text book we can write system~(\ref{Eq:al1}) as 
\begin{mathletters}
\label{Eq:al7}
\begin{eqnarray}
\Box\, \bA &=& -\frac{4\pi}{c} \bj_{\rm tot} = -\frac{4\pi}{c}
\bigl[\bj_{\rm free} + \frac{\p \bP}{\p t} + c\, {\rm curl}\, \bM \bigr] 
\label{E7a}\\
\Box \phi &=& -4\pi \bigl[\rho -{\rm div}\,\bP \bigr] \label{E7b}
\end{eqnarray}
\end{mathletters}
under Lorentz gauge, i.e., 
${\rm div}\, \bA + (1/c) (\p \phi/\p t) = 0$ and
\begin{mathletters}
\label{Eq:al9}
\begin{eqnarray}
\Box \bA &=& -\frac{4\pi}{c} \bigl[\bj_{\rm tot} - \frac{1}{4 \pi} 
\nabla \frac{\p \phi}{\p t}\bigr] \label{E9a}\\
\nabla^2 \phi &=& -4\pi \bigl[\rho -{\rm div}\bP \bigr] \label{E9b}
\end{eqnarray}
\end{mathletters}
under Coulomb gauge, i.e., ${\rm div} \bA = 0$.
Here $\Box = \nabla^2 - (1/c^2) (\p^2 /\p t^2).$ 
Note that to obtain~(\ref{Eq:al7}) or~(\ref{Eq:al9}) it is sufficient to 
consider~(\ref{E1a}) and~(\ref{E1b}) only since the two others are fulfilled 
identically. 
Let us now develop two potential formalism. Two potential formalism was 
first introduced in~\cite{DMag} and further developed in~\cite{Kluwer,ICTP}. 
In both papers we introduce only two vector potentials 
$\bal^m,\,\bal^e$ and use only the curl-subsystem of the Maxwell equations 
with the additional condition ${\rm  div} \bal^{m, e} = 0$. 
Thus, in our view our previous version of two potential formalism lack of 
completeness. In the present paper together with the vector potentials 
$\bal^m, \bal^e$ we introduce two scalar potentials $\vf^m$ and 
$\vf^e$ such that 
\begin{mathletters}
\begin{eqnarray}
\bbe &=&
{\rm curl}\bal^m + \frac{1}{c} \frac{\p \bal^e}{\p t} + \nabla \vf^m, \\
\bde &=&
\mbox{curl}\bal^e - \frac{1}{c}\frac{\p \bal^m}{\p t} - \nabla \vf^e 
\end{eqnarray}
\end{mathletters}
It can be easily verified that system of equations~(\ref{New}) 
are invariant under this transformation and take the form 
\begin{mathletters}
\label{LG}
\begin{eqnarray}
\Box\, \bal^m &=& - \frac{4 \pi}{c}\bigl[\bj + c\, {\rm curl}\, \bM + 
c\, {\rm curl\,curl}\,\bT^m\bigr], \\ 
\Box\, \vf^m &=& 0 \\
\Box\, \bal^e &=& -\frac{4 \pi}{c}\bigl[{\rm curl}\, \bP +  
{\rm curl\,curl}\,\bT^e\bigr], \\ 
\Box\, \vf^e &=& - 4 \pi\, \rho 
\end{eqnarray}
\end{mathletters}
under 
${\rm div}\,\bal^{m,e} + (1/c) (\p \vf^{e,m}/\p t) = 0$
and 
\begin{mathletters}
\label{CG}
\begin{eqnarray}
\Box\, \bal^m &=& - \frac{4 \pi}{c}\bigl[\bj + c\, {\rm curl}\, \bM + 
c\, {\rm curl\,curl}\,\bT^m - \frac{1}{4\pi} \nabla 
\frac{\p \vf^e}{\p t}\bigr] \label{E14a}\\ 
\nabla^2\, \vf^m &=& 0 \label{E14b}\\
\Box\, \bal^e &=& -\frac{4 \pi}{c}\bigl[{\rm curl}\, \bP +  
{\rm curl\,curl}\,\bT^e -\frac{1}{4\pi}\frac{\p \vf^m}{\p t} \bigr]
\label{E14c}\\ 
\nabla^2\, \vf^e &=& - 4 \pi\, \rho \label{E14d}
\end{eqnarray}
\end{mathletters}
under ${\rm div}\, \bal^{m,e} = 0.$
The solutions to the systems~(\ref{LG}) and~(\ref{CG}) can be written as
follows (see for example~\cite{ICTP,ODJ}):
The solutions to the d'Alembert equation
\begin{equation}
\Box F (\br, t)= f(\br, t)
\label{Alem}
\end{equation}
look
\begin{equation}                  
F (\br, t) = -\frac{1}{4\pi} \int\limits_{\mbox{all space}} \frac{f(\br', t') 
d\br'}{|\br - \br'|}\Biggr|_{t' = t - |\br - \br'|/ c} 
\label{}
\end{equation}
whereas the solutions to the Poisson equation 
\begin{equation}
\nabla^2 F(\br) = f(\br) 
\label{Pois}
\end{equation}
read
\begin{equation}
F (\br) = -\frac{1}{4\pi} \int \frac{f(\br') d\br'}
{|\br - \br'|}
\label{}
\end{equation}
It is necessary to emphasize that the potential descriptions 
electrotoroidic and magnetotoroidic media are completely separated. The 
properties of the magnetic and electric potentials $\bal^m$ and $\bal^e$ 
under the temporal and spatial inversions are opposite \cite{DK}. 
The potential $\bal^e$ ($\bal^m$) is related to the toroidness of the
medium $\bT^e$ ($\bT^m$) as $\bB$ ($\bD$) to $\bM$ ($\bP$). 
Note that if $\mbox{div}\bde \ne 0$ and there does exist free current in 
the medium we have to use the direct method for finding all constrains in 
the theory suggested by Dirac. Dirac applied his method to electrodynamics 
and found that electromagnetic potentials have only two degrees of freedom 
described by transverse components of vector potential. This method was 
developed by Dobovik and Shabanov~\cite{DSh}, where classical and quantum 
dynamics of a system of non-relativistic charged particles were considered.
  
\section{Conclusion}
\setcounter{equation}{0}
The modified equations of electrodynamics has been obtained in account
of toroid moment contributions. The two-potential formalism has been
further developed for the equations obtained. Note that introduction
of free magnetic current $\bj_{\rm free}^{m}$ and magnetic charge
$\rho^{m}$ in the equations~(\ref{N3}) and ~(\ref{N4}) respectively
leads to the equations obtained by Singleton~\cite{Singleton}.

\end{document}